\documentclass[12pt,preprint]{aastex}
\usepackage{psfig}
\usepackage{natbib}

\newcommand{\sgr}{SGR~0526$-$66}
\newcommand{\rxj}{\objectname{RX~J052600.3$-$660433}}
\newcommand{\errrad}{{\ensuremath{2\farcs3}}}

\newcommand{\eg}{{e.g.\ }}

\newcommand{\lsim}{\lesssim}
\newcommand{\gsim}{\gtrsim}
\begin{document}
\shorttitle{HST Observations of {SGR~0526$-$66}}
\shortauthors{Kaplan {et al.}}

\title{HST Observations of {SGR~0526$-$66}: New Constraints
on Accretion and Magnetar Models} 
\author{D.~L.~Kaplan\altaffilmark{1}, S.~R.~Kulkarni\altaffilmark{1},
M.~H.~van~Kerkwijk\altaffilmark{2}, R.~E.~Rothschild\altaffilmark{3},
R.~L.~Lingenfelter\altaffilmark{3}, D.~Marsden\altaffilmark{4},
R.~Danner\altaffilmark{5}, \& T.~Murakami\altaffilmark{6}}
\altaffiltext{1}{Department of Astronomy, 105-24 California Institute of
Technology, Pasadena, California, 91125, USA}
\altaffiltext{2}{Sterrenkundig Instituut, Universiteit Utrecht,
Postbus 80000, 3508 TA Utrecht, The Netherlands} 
\altaffiltext{3}{Center for Astrophysics and Space Sciences 0111,
University of California, San Diego, La Jolla, California 92093-0111, USA}
\altaffiltext{4}{NASA Goddard Space Flight Center, Mailstop 662.0,
Greenbelt, MD 20771, USA}
\altaffiltext{5}{Jet Propulsion Laboratory, California Institute of Technology,
4800 Oak Grove Drive, Pasadena, California 91109-8099, USA} 
\altaffiltext{6}{ISAS, 3-1-1 Yoshinodai, Sagamihara, Kanagawa, Japan 229.}
\email{dlk@astro.caltech.edu}

\slugcomment{Accepted by ApJ}

\begin{abstract}
Soft $\gamma$-ray Repeaters (SGRs) are among the most enigmatic sources known
today.  Exhibiting huge X- and $\gamma$-ray bursts and flares, as well as
soft quiescent X-ray emission, their energy source remains a mystery.
Just as mysterious are the Anomalous X-ray pulsars (AXPs), which share many
of the same characteristics.  Thanks to recent {\it Chandra}
observations, \sgr, the first SGR, now appears to be a
transition object bridging the two classes, and therefore observations
of it have implications for both SGRs and AXPs.  The two most popular
current models for their persistent emission are accretion of a fossil disk or
decay of an enormous ($\sim 10^{15}$~G) magnetic field in a magnetar.
We show how deep optical observations of \sgr, the only SGR with small
enough optical extinction for meaningful observations, show no
evidence of an optical counterpart. These observation place strong new
constraints on both accretion disk and magnetar models, and suggest
that the spectral energy distribution may peak in the hard-UV.
Almost all accretion disks are excluded by the optical data, and 
a magnetar would require a $\sim 10^{15}$--$10^{16}$~G field.
\end{abstract}


\section{Introduction}
Soft $\gamma$-ray Repeaters (SGRs; for a recent
observational review, see \citealt{h99}) were initially discovered by
their intense and repeated 
emission of soft ($kT \leq 30$~keV) $\gamma$-rays.  These bursts are
significantly super-Eddington ($\gsim 10^{3} L_{\rm
Edd}$; \citealt{h99}), but are dwarfed by the giant flares.  The flares are 
thousands of times more energetic than 
typical bursts and have  harder spectra ($\sim$~MeV), but are rare,
with only two observed in the past 20 years. 

SGRs also emit quiescently in X-rays.  This emission exhibits a
power-law spectrum with photon index $\Gamma\sim 2$, and may also
have small blackbody contributions with $kT\approx 0.5$~keV.  SGRs
pulsate both in quiescence and during bursts with periods (5--8~s)
that are longer  than those typical for radio pulsars, and relatively
large spin-down rates as well  ($\dot{P} \gsim 10^{-11}\mbox{ s
s}^{-1}$; \citealt{kds+98,h99}).

Based on energetics and
proximity to supernova remnants (SNRs) or star-formation
regions, 
SGRs are generally thought to be young
($\lsim 10^{4}$~yr) neutron stars.  
Given this, the most widely accepted model for SGRs has been the
magnetar model 
\citep{td93}. 
Magnetars, 
neutron stars with $B\gsim 10^{15}$~G, are objects whose
primary power source is magnetic field decay rather than
spin-down (radio pulsars) or accretion (X-ray binaries).  The high
magnetic fields were  invoked 
to explain the very long super-Eddington tail of the 5~March~1979
outburst \citep{p92} and 
its spin period \citep{dt92}.
In the magnetar model, SGRs produce bursts and
flares through violent recombination and unpinning 
of the magnetic field, driven by crust fractures \citep{td95}.
Non-thermal quiescent emission is thought to be  due to
currents in their  magnetospheres  arising from the continuing
competition between crust and 
magnetic stresses \citep{td96,t00}.

The Anomalous X-ray
Pulsars (AXPs; see \citealt{m99} for a review) are a group of objects
that appear similar to the SGRs. However, they have softer spectra
($\Gamma\sim 4$
and more prominent blackbody components) and have not been observed to 
burst. \citet{td96} proposed 
that both classes had similar origins, based on similar  spin
properties and X-ray luminosities.

While intriguing, magnetars are not the only possible models for AXPs.
Propeller driven spindown of accreted/ejected matter
from a fallback disk (\citealt*{vptvdh95}; \citealt*{chn00})
produced during the supernova explosion, a ``pushback'' disk
(enhanced fallback from an overly dense environment;
see \citealt{mlrh99,mlrh00}), or a disk acquired by a high velocity
neutron star that has caught up with its ejecta \cite{mlrh99}, was
postulated   
to explain the emission and period clustering
of the AXPs.  With a standard $10^{12}$-G magnetic field, an accreting
pulsar would naturally reach an equilibrium spin period of $\sim 8$~s, and
the observed variations in spin-down (\citealt{kgc+00};
\citealt*{klc00}) could be explained by accretion torques.

Accretion would provide a large and easily accessible
reservoir of energy for the AXPs.  This is in contrast to magnetars,
which can only emit $E \sim
B^{2}/8\pi \times 4\pi R^3/3=10^{47}\mbox{ ergs}$ (for a $10^{15}$-G
field) over their lifetimes.
Given the attractiveness of the accretion model on energetics ground,
it has also  been proposed that SGRs are powered by the same mechanism
\citep{mlrh99}.  
In the propeller spin-down model, the bursts are 
expected to result  
from starquakes produced \citep[e.g.][]{rbc+80,rbl81,ek83},
by fractures 
and phase transitions in subducted crust piled up by the extreme plate
tectonics \citep{r91} driven by the very rapid spindown of the 
neutron star. The damping times of neutron star vibrational modes in 
such a quake model can possibly explain \citep{rbl81} both the duration 
of the 5~March~1979 outburst and its very long super-Eddington tail.

The discovery of the soft  power-law 
spectrum ($\Gamma=3.24$) of one SGR, \sgr\, by
\citet[hereafter K00]{k+00}  suggests AXPs and SGRs may be linked through
spectral as well as spin properties, and that \sgr\ may be a
transition object bridging the two classes.

With 
its high Galactic latitude, \sgr\ has a comparatively low extinction,
allowing much deeper optical 
observations than those of the other SGRs, despite  its greater distance.
The other SGRs have extinctions
of 10--30 magnitudes  in the optical  
band ($A_{V}\approx 13$ for SGR~1900$+$14 based on \citealt{hlk+99};
$A_{V}\sim 28$~mag for SGR~1806$-$20; \citealt{vkkmn95}).  Therefore,
this object  
provides an important test for models of the quiescent optical 
emission from SGRs.

\section{SGR~0526$-$66}
\label{sec:sgr}
On 5~March~1979, spacecraft recorded a giant $\gamma$-ray flare
\citep{mgi+79,cdp+80} that is currently 
second in brightness, behind only  the 28~August~1998 flare of
SGR~1900$+$14 \citep{hcm+99}.
 The burst showed a very short rise ($< 1$~ms) followed by a $\sim
 150$-ms decay and 
a pulsating phase (lasting $>2$~min; \citealt{cdp+80}).   While overall
the spectrum is rather soft, the initial flare had a hard tail,
perhaps associated with the initial outburst rather than its
afterglow \citep{mgi+79}.  There 
were also indications of  a 430-keV emission line, probably the 511-keV
annihilation line redshifted by gravity from the
surface of a neutron 
star \citep{mgi+79}.  \citet{bch+79} and \citet{mgi+79} identified an
8-s periodicity in the burst data.

Triangulation located the source in the Large Magellanic Cloud (LMC),
specifically in the direction of SNR \objectname{0525$-$66.1} 
\citep[N49;][]{ekl+80}. Assuming the association to hold, the inferred
isotropic luminosity of the 
source was calculated to be  $5 \times 10^{44}\mbox{ erg s}^{-1}=2
\times 10^{6} L_{\rm Edd}$, with a total energy output of $\sim
7 \times 10^{44}\mbox{ erg}$ (for X-rays $>50$~keV;
\citealt{mgi+79,h99}).  

On 6~March~1979 a second burst
lasting $\sim 1.5$~s was observed at the same position, with a similarly
soft spectrum, although its intensity was a factor of 100 below that
of 5~March~1979.  Bursts of lower intensity were subsequently
observed \citep{rl84}, although none after $1983$.  The later
bursts showed a clustering which was interpreted  
\citep{rl84} as a periastron passage in a highly eccentric 164 day 
orbit, which was later also indicated in the observation of and 
optical flash from \sgr\ \citep{pdh+84}.  However, later analysis of
the clustering of bursts from SGR~1900$+$14 which has also emitted a
giant flare) and SGR~1806$-$20 has pointed towards a lognormal
distribution of time intervals, similar to that of earthquakes
\citep{ceg+96,gwk+99,gwk+00}.  This distribution suggests that the
bursts result from internal rather than external triggers.

Later, X-ray observations identified  
a point source, \rxj, coincident with the $\gamma$-ray error
box \citep*{rkl94,mrlp96}.  The luminosity of the quiescent source is $\sim
10^{36}\mbox{ erg s}^{-1}$ (0.1--2.4~keV; \citealt{mrlp96}; K00), and 
while no periodicity was found in the ROSAT data, the limit on the pulsed
fraction of 66\% \citep{mrlp96} was not very stringent.   Recent {\it
Chandra X-ray Observatory} (CXO) observations of \rxj\
have shown that the source is indeed pointlike and has a drastically
different spectrum (nonthermal with photon index of $3.24$) than the
rest of the 
supernova remnant, leaving little doubt that \rxj\ is the quiescent
counterpart of \sgr\ (K00).

\subsection{Distance and Reddening to SGR~0526$-$66}
\label{sec:red}
To find the extinction $A_{V}$, \citet{mrlp96} use a relation
between the hydrogen column density $N_{H}$ ($8.3 \times 10^{21}\mbox{
cm}^{-2}$; K00) and the extinction.  Most of the 
intervening hydrogen and extinction are local to the LMC
($A_{V,{\rm Galactic}}\approx 0.1$, as inferred
from \citealt{bh82}), so we therefore use the 
relation appropriate to the LMC: $A_{V}=N_{H}/8.3\times
10^{21}\mbox{ cm}^{-2}$ \citep{wd00}
and find $A_{V}=1.0$~mag.    We then use the standard Galactic reddening
curve (which 
should also apply to the LMC longward of 200~nm; \citealt{nmw+80,z99})
from \citet{s77}  
normalized to $A_{V}$.  From this we find 
the reddening $E_{B-V}=0.3$~mag, generally consistent with that
found by \citet{vblr92} from studies of line ratios in N49 (although
there is significant variation over 
the remnant,
with $E_{B-V}$ ranging from 0.0~mag to 0.5~mag).  
We note that the values for the extinction are somewhat
imprecise and may 
have significant variations on scales as small as $2\arcsec$
\citep{vblr92}, but this does not significantly affect our analysis.
The extinction values that we used are listed 
in Table~\ref{tab:sum}, and are at the upper end of the range possible
for \sgr.  

We parameterize the distance
to the LMC/\sgr\ as $D_{\rm LMC}=50d_{50}$~kpc.

\section{Observations}
The data consist of {\it Hubble Space Telescope} (HST) WFPC2
observations with several filters of 
the field centered 
around the supernova remnant N49 in the LMC.  See Table 
\ref{tab:sum} for a log of the observations.  
The HST WFPC2 filter designations describe the center wavelengths and
widths.  F300W, for example, is a wide bandwidth filter centered at
300~nm, while 
F547M is a medium bandwidth filter centered at 547~nm.

\subsection{Image Reduction}

The data from 14~November~1998 and 27~April~1999 were taken with HST in
different orientations, so they were processed separately, although
with the same procedure.  

We used the drizzling\footnote{Drizzling is a technique
to combine HST images from slightly different positions and
orientations into a single image.  The HST PC has
$0\farcs046$~pixels, and the nominal resolution
is $1.2\lambda/D=0\farcs05$, indicating that it undersamples the
point-spread function (PSF).   By dithering to positions with
fractional-pixel offsets, we are able
to retrieve some of the lost information, eventually
increasing 
the nominal resolution by a factor of $\sqrt{2}$, while preserving
photometric integrity.} procedure to reduce the images and remove
cosmic rays \citep{fm98},
employing the standard methodology and parameter values.
We combined the
resulting images into a master $2048\times 2048$~ pixel image,
oversampling the Planetary Camera (PC) pixels 
by a factor of two (changing the pixel scale from $0\farcs046 \mbox{
pixel}^{-1}$ to $0\farcs023 \mbox{ pixel}^{-1}$).  We also
applied  geometric distortion corrections appropriate for each
wavelength, so that the relative astrometry between detectors or in a
given detector should be accurate.
Typical stellar sources had a FWHM of $2.8\mbox{ pixels}=0\farcs06$,
as expected from the telescope PSF.

\subsection{Astrometry}
\label{sec:astrom}
\begin{figure*}
\epsscale{0.5}
\centerline{
\vbox{\hbox{
\plotone{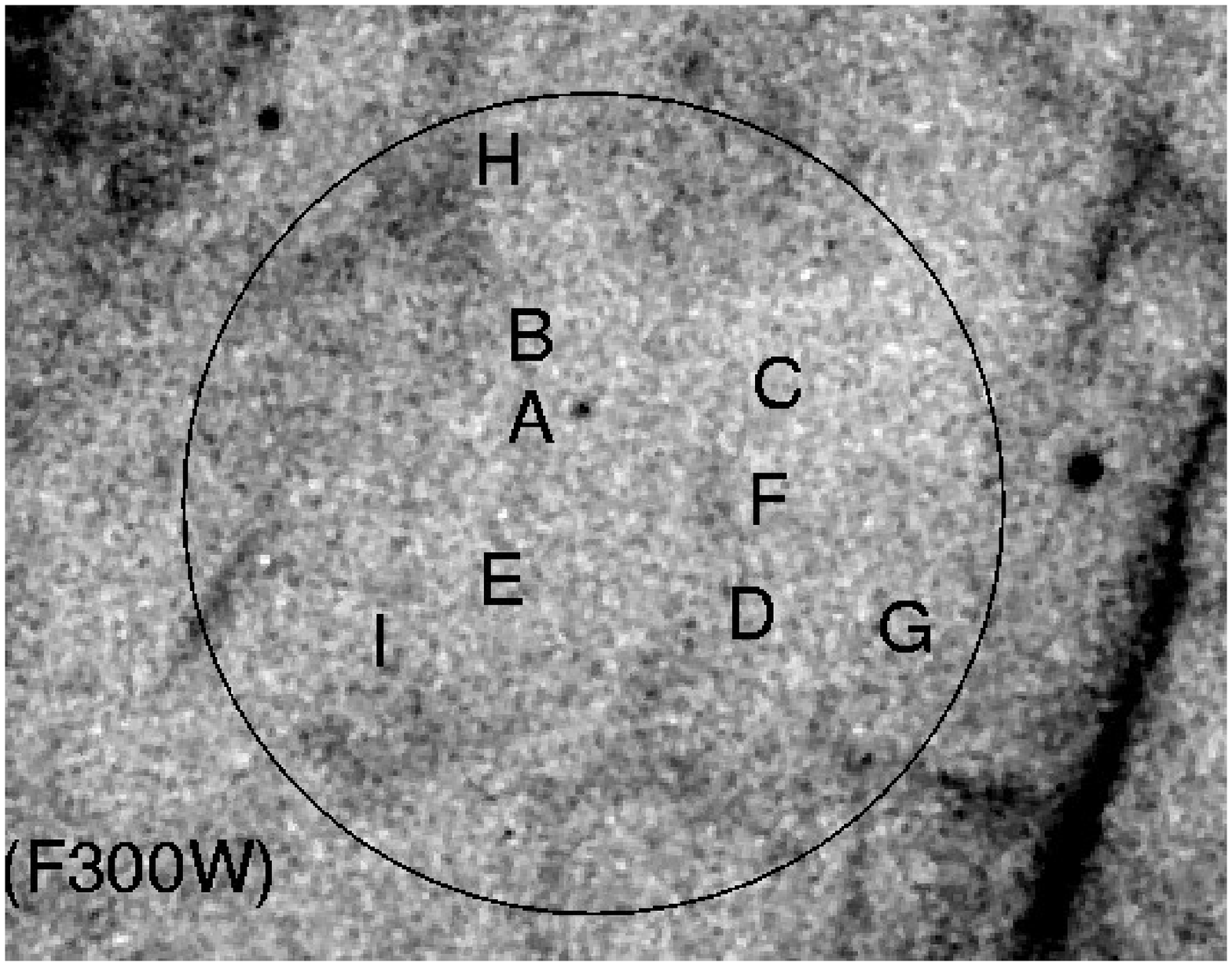}
\plotone{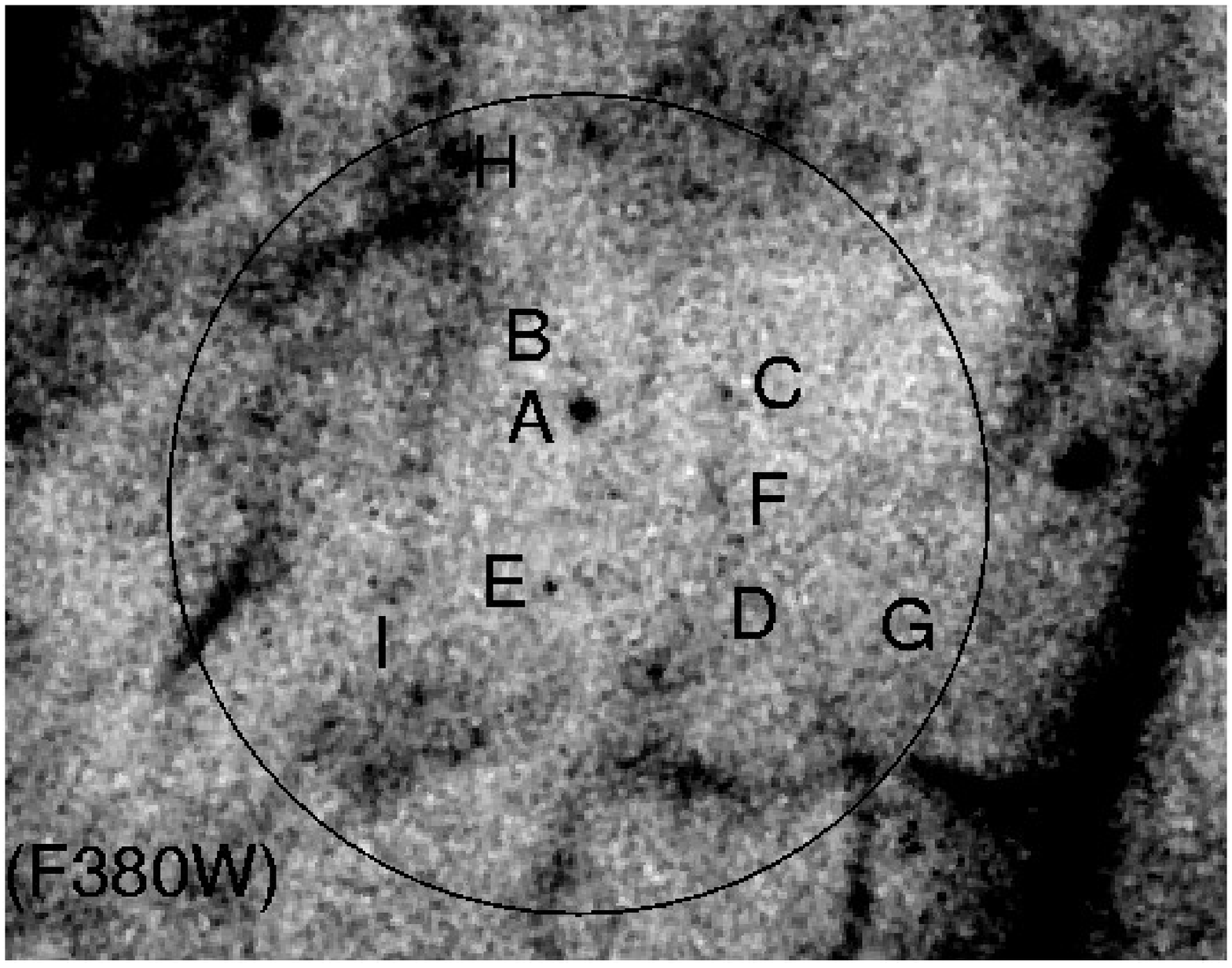}}
\hbox{
\plotone{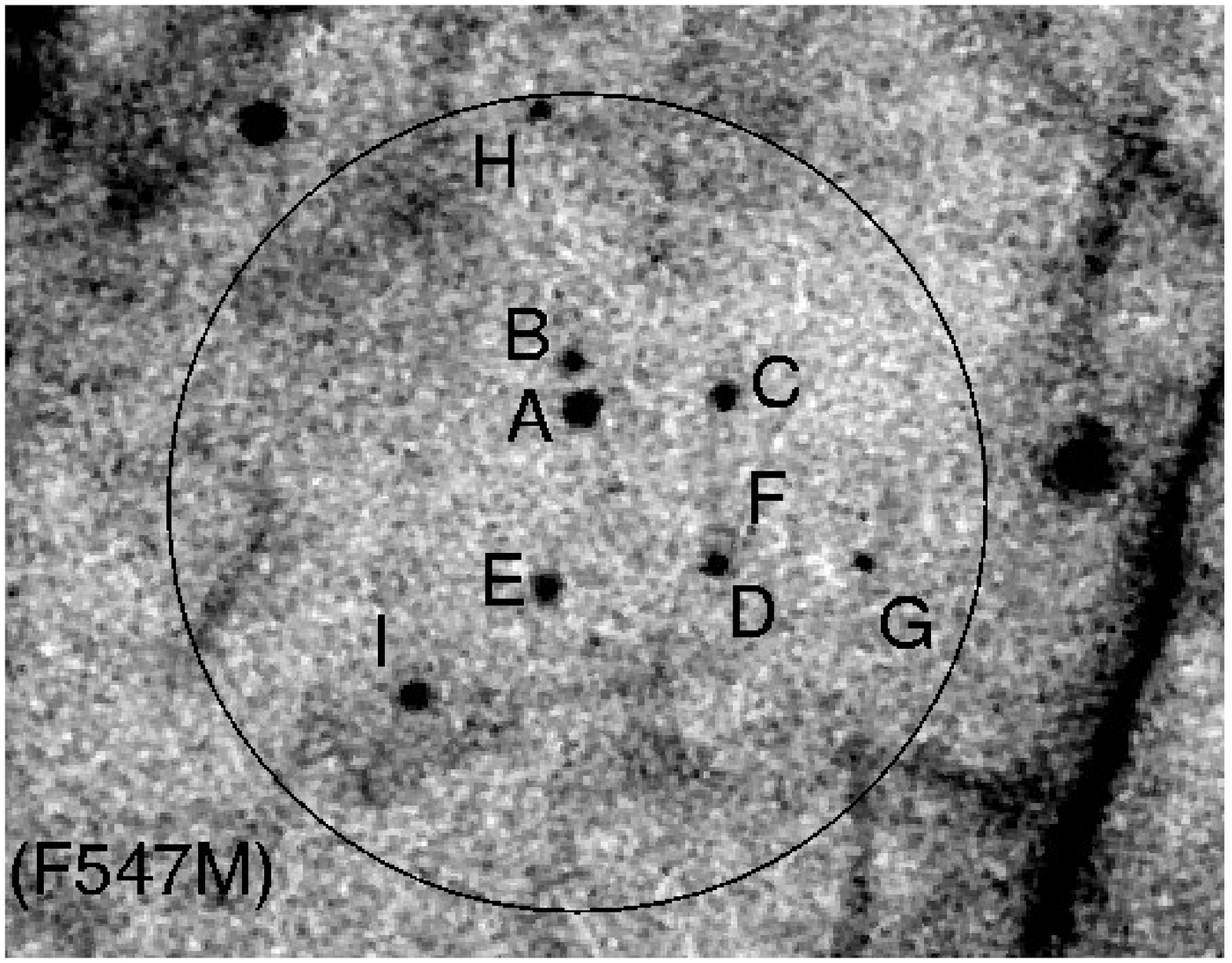}
\plotone{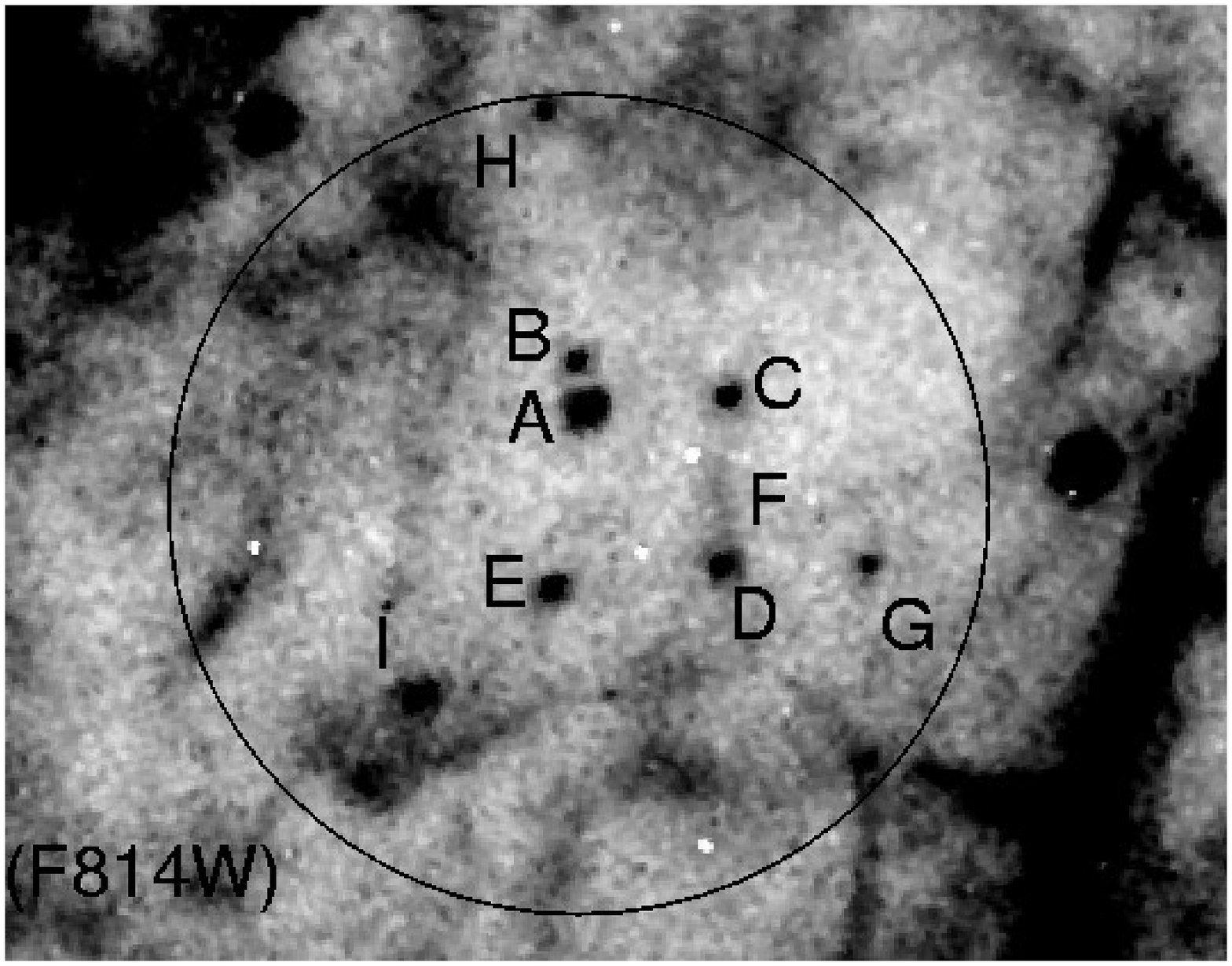}}}}
\figcaption[figure1a.eps,figure1b.eps,figure1c.eps,figure1d.eps]{Images of the region
around  \sgr.  Filters are F300W (upper left), 
F380W (upper right), F547M (lower left), and F814W (lower right).  A
\errrad\ radius circle is 
indicated, as are the sources 
described in Section~\ref{sec:anal} and Table~\ref{tab:sources}.
North is up and East is to the left.\label{fig:circle}}
\end{figure*}

We used the USNO-A2.0 catalog \citep{m98} as our astrometric reference,
fitting to the F547M images (which presented the best combination of
signal to noise ratio and nebular rejection).
While no USNO-A2.0 stars were visible on the PC image, we had many on
the Wide Field (WF) images.  Using twenty-four USNO-A2.0 stars, we
computed a solution for a mosaiced image of all the 
WFPC2 detectors assembled with the task {\tt wmosaic}.  The {\tt
wmosaic} task corrects for geometric distortions and relative offsets
between the different detectors, so using the mosaiced image to solve
for stars on an individual detector should not have introduced significant errors.
The
solution has $0\farcs3$ rms residuals in each coordinate (comparable to
the accuracy of the USNO-A2.0 positions; \citealt{m98}).  We then
identified fourteen fainter stars on the PC portion of the 
mosaiced image, determined their positions from the mosaic plate solution,
and found the solution for the drizzled images ($0\farcs03$ residuals for each
coordinate), which we  
transferred to the other bands.   The
position for \rxj\ [determined by K00 
to be
$\alpha(J2000)=05^{\rm h}26^{\rm m}00\fs819$,
$\delta(J2000)=-66\deg04\min36.48\sec$] was located on the drizzled images,
yielding the position  in Figure~\ref{fig:circle}.   The final intrinsic
astrometric uncertainties are $0\farcs4$, and the CXO systematic
uncertainties are $1\farcs0$ (based on other
data, we believe $1\farcs0$ to be a conservative estimate for CXO;
E.~Schegel~2000, private communication).
This gives $1\farcs1$ for the overall 1~$\sigma$ uncertainty in each
coordinate.  From this we find the 90\% confidence radius to be \errrad.

\subsection{Photometry}
\label{sec:phot}
The position of \rxj, as measured with CXO by K00, is on the PC
detector, so we are only concerned with these data.

Once we had drizzled images for each filter, we 
performed aperture photometry on each object on the PC using the {\tt
SExtractor} photometry 
package \citep{ba96}. 
Using the F547M image as the reference, we measured the positions of
$220$ stars on the PC.   We then used this
master list of sources to do 
photometry for all of the filters.  Because of the large amount of
nebulosity in the images, it is difficult to determine the 
background level accurately.  We computed a background by gridding the
image into $32\times 32$ pixel 
 bins ($=0\farcs7 \times 0\farcs7$),
and then producing an average (after rejecting high pixels) for each
one.  We subtracted this smooth background, resulting in an image that
does not contain most of the variations due to nebulosity.   Next, a  
fitted elliptical profile was analyzed for each source.  The magnitude
was determined by fitting this profile out to a radius of $2\times \mbox{FWHM}=0\farcs12$
(for stars) and integrating.   We then applied two types of
aperture-corrected magnitude estimates with {\tt SExtractor} (``adaptive
aperture'' and ``corrected isophotal'' magnitudes) to compute the true
magnitude.  

Zero-point
magnitudes in the STMAG system\footnote{All magnitudes are in the STMAG system, where
$m=-21.1-2.5 \log F_{\lambda}$, with $F_{\lambda}$ in $\mbox{ ergs
s}^{-1}\mbox{ cm}^{-1}\mbox{ \AA}^{-1}$.} were calculated according to
the WFPC2 
manual\footnote{\url{http://www.stsci.edu/instruments/wfpc2/Wfpc2\_phot/wfpc2\_cookbook.html}}, 
where the calibration parameters  {\tt PHOTFLAM} and {\tt PHOTZPT} are
taken from the image headers.  The 
values we calculated are listed in Table~\ref{tab:sum}.  To convert
the magnitudes to fluxes, we scale our count rates 
by the {\tt PHOTFLAM} parameter, which is the flux of a source that produces
$1\mbox{ DN s}^{-1}$.    

To determine limiting magnitudes, we  empirically determined the
point-spread functions (PSFs) of the 
different images for $\sim 30$ stars.  We then used this PSF to add 50
simulated 
stars to the central part of the images.  These stars were distributed
in magnitude such that some were above and some below the detection
limit.  We performed photometry on these simulated stars, and 
define the limiting (3$\sigma$) magnitude to be where the measured
magnitudes of 
the stars have errors $\gsim 0.3$~mag, indicating that stars can
no longer be accurately 
measured.  These values have an intrinsic uncertainty of $\sim
0.1$~mag, and are 
reported in Table~\ref{tab:sum}.  We note that these values represent
completeness 
limits for the region around \sgr\ --- individual stars can still be
detected with fainter magnitudes, but no source was missed to the limits
presented here.

\section{Analysis}
\label{sec:anal}
Our primary goal for these observations was to detect a counterpart to
\sgr\ or a cooling synchrotron nebula
left by previous bursts (cf.\ the radio nebula from \citealt*{fkb99}) and
confined by the hot gas in the interior of N49.
We therefore examined all of the sources within the
 \errrad\ error circle that were not obviously part of the SNR.  The images
are shown in 
Figure~\ref{fig:circle}.  We present  photometric
data concerning the sources in the error circle in
Table~\ref{tab:sources}.  

\begin{figure*}
\epsscale{0.5}
\centerline{
\vbox{
\hbox{
\plotone{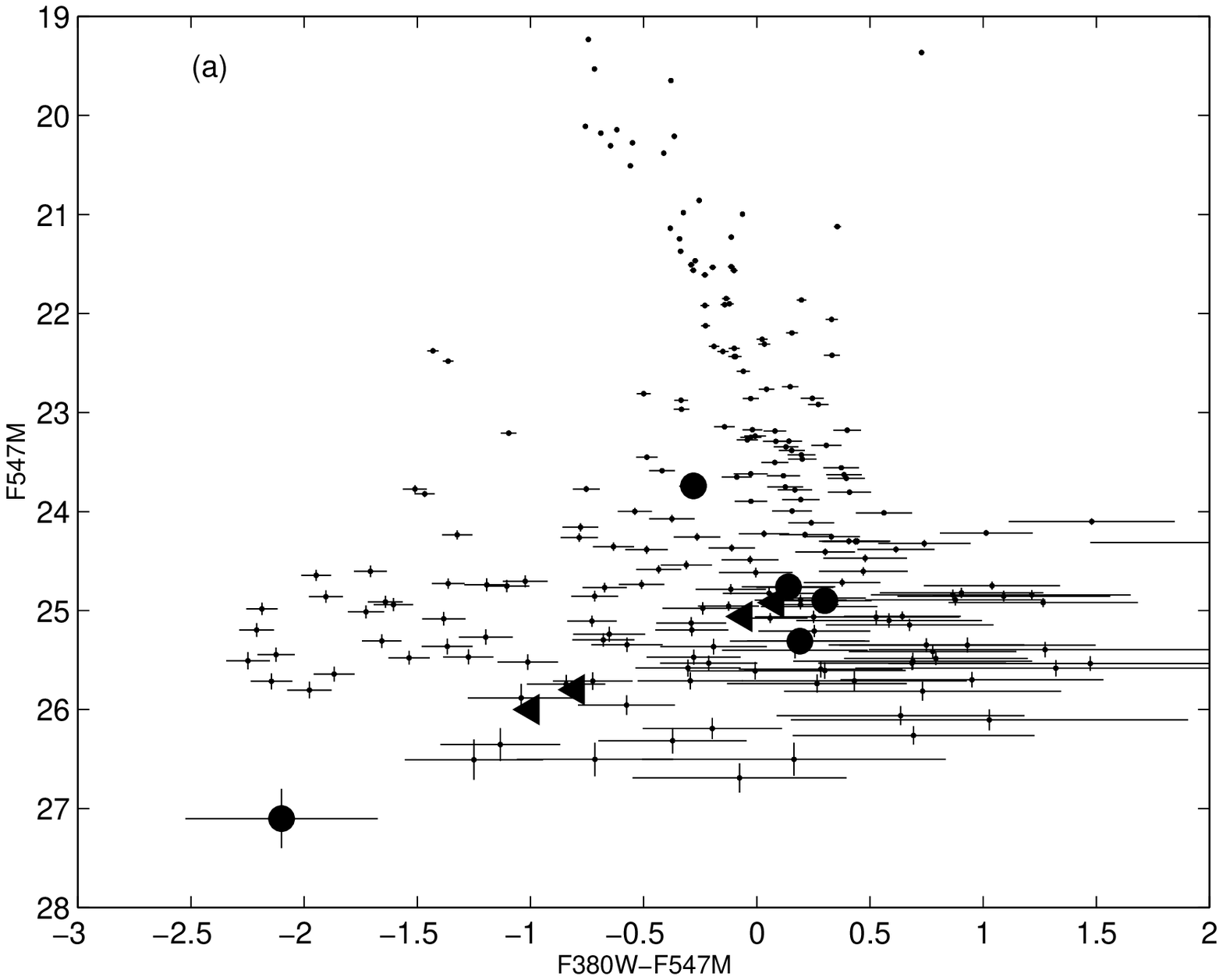}
\plotone{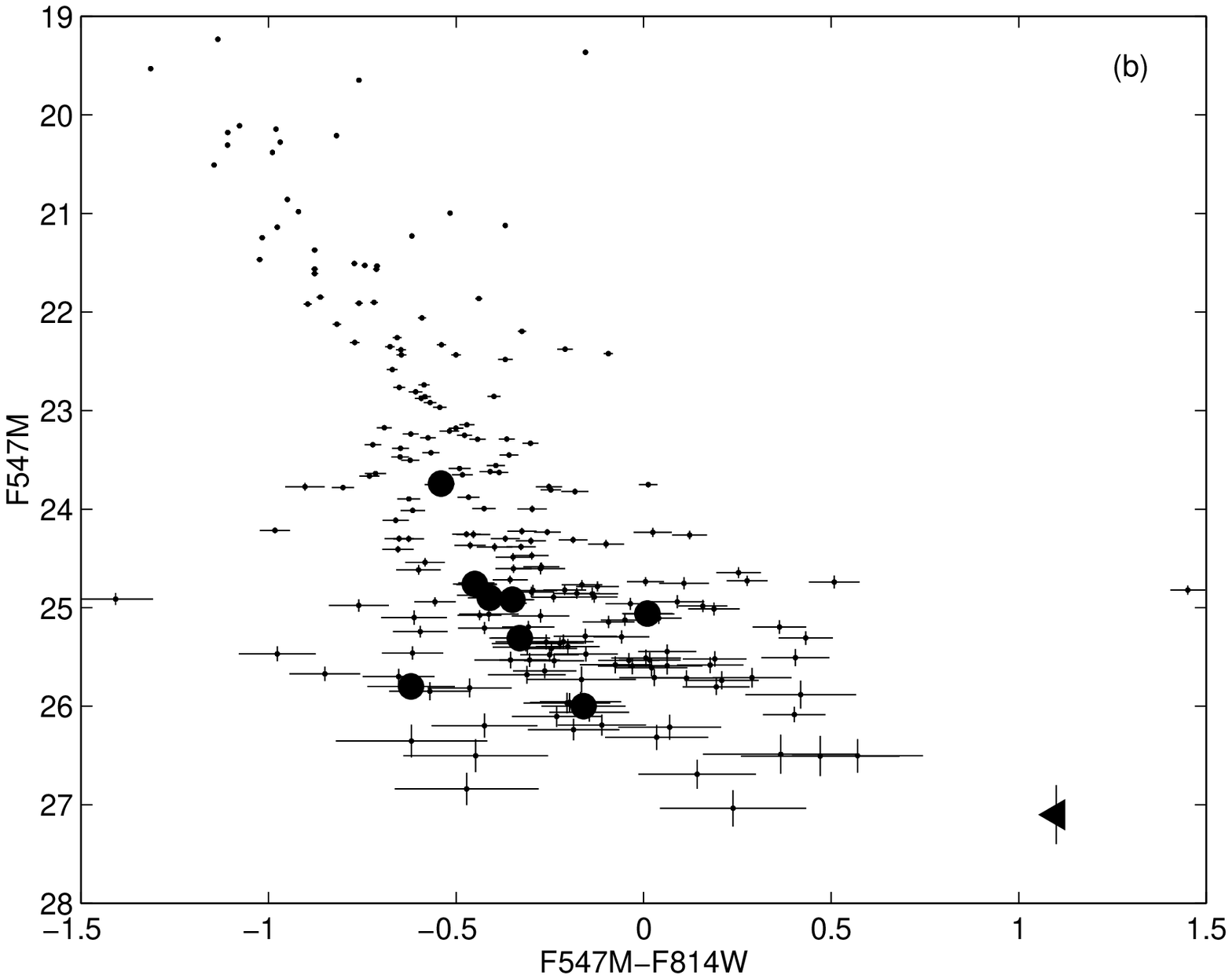}}
\hbox{
\plotone{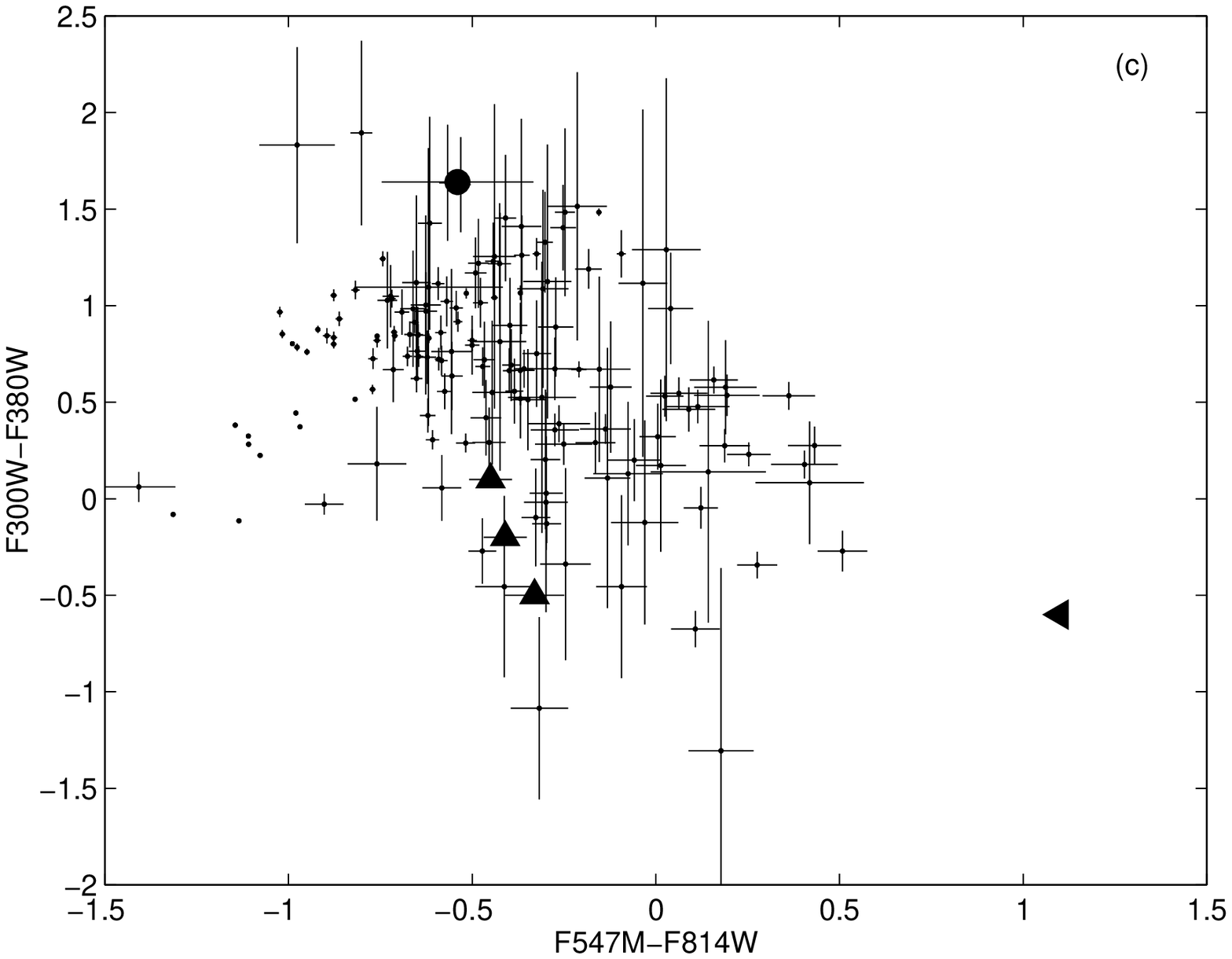}
\plotone{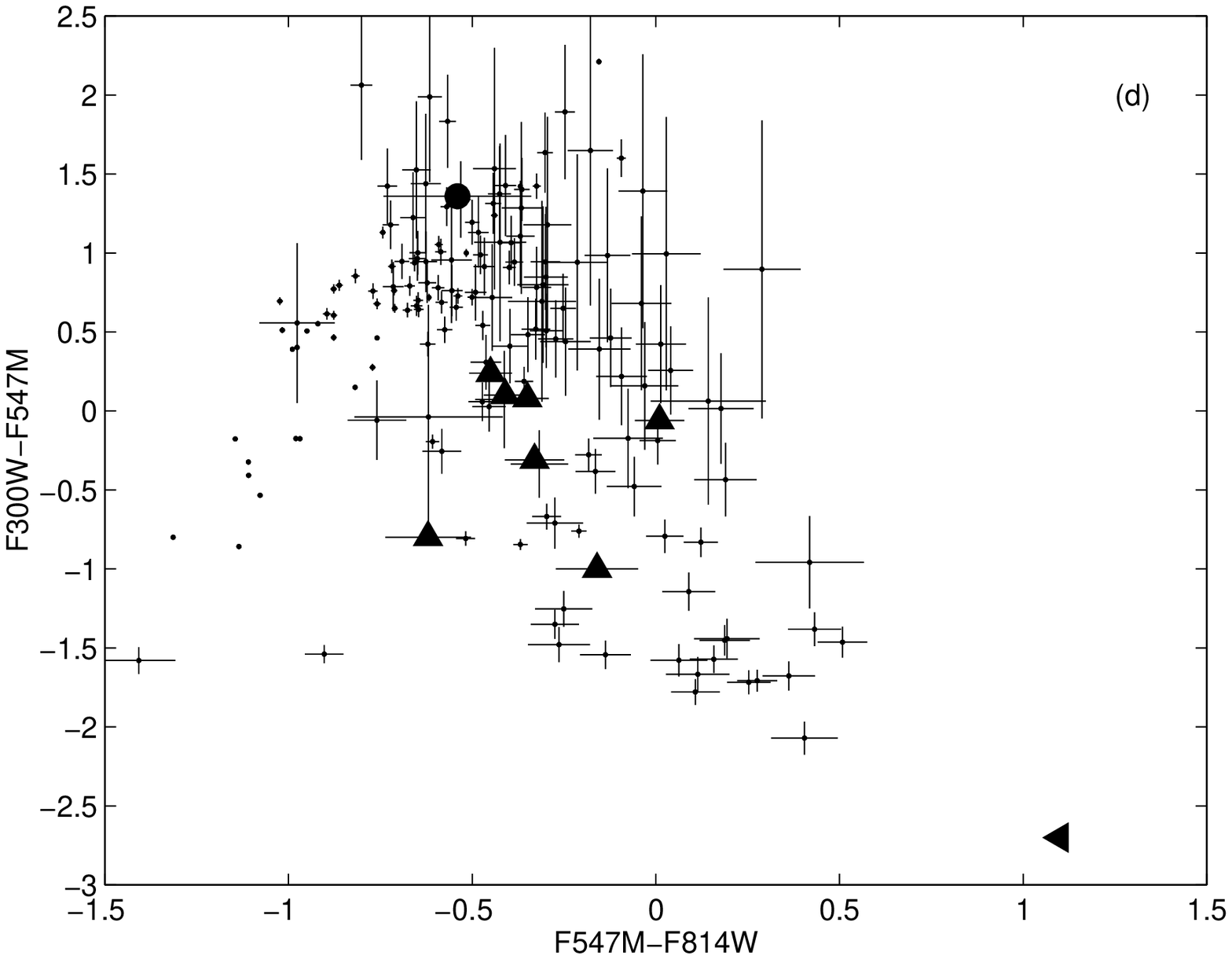}}}}
\figcaption[figure2a.eps,figure2b.eps,figure2c.eps,figure2d.eps]{(a): F547M magnitude vs.\
F380W$-$F547M color 
for all of the detectable 
sources (220) in the field.  (b): F547M magnitude vs.\ 
F547M$-$F814M color 
for same sources.  (c): F300W$-$F380W color vs.\ F547M$-$F814W
color.  (d):  F300W$-$F547M color vs.\ F547M$-$F814W
color.  Magnitudes are not corrected for
reddening.  Error-bars represent $1\sigma$.  The large,
solid circles/arrows are those 
listed in Table \ref{tab:sources}; the labels are indicated in
Figure~\ref{fig:hr2}.  The main sequence  
is clearly visible.\label{fig:hr1}}
\end{figure*}

We compared the colors of these sources  to those of
the $\sim 200$ other sources detectable on the PC image.  All but one
of the nine candidate sources lie 
in the color-magnitude diagram formed by the field sources, as seen in
Figure~\ref{fig:hr1}. 

Seven (possibly eight) of the sources from Table~\ref{tab:sources} are
generally consistent with main sequence G/K 
stars at the distance of the LMC, with reddening similar to that used
here ($E_{B-V} \lsim 0.3$).  Of the two that are not (F and possibly
E), both have nebulosity 
leading to other  
parts of the SNR.  We now examine these sources in detail.
For source E we find $M_{\rm F380W} \approx 6.2$, $M_{\rm F547M}
\approx 5.4$, and $M_{\rm F814W}
\approx 6.2$.  This may be consistent with a GV star (given reddening
uncertainties), but the source 
has some surrounding nebulosity trailing to the South and West, in the
direction of other filaments.  We therefore believe that this is part
of the SNR.
For source F, we find $M_{\rm F300W} \approx 4.1$, $M_{\rm F380W} \approx
5.0$, and $M_{\rm F547M} \approx 7.6$, and a slightly extended
morphology.  This is not consistent with a 
  main-sequence star.  However, the source is extended towards the South
 thereby making the 
photometry suspect,  and the source merges into the
SNR nebulosity.  In the F814W image F appears almost to have a
bow-shock morphology, but this morphology changes so much in the other
images that we believe it to be coincidence.  We therefore conclude
that F, like E, is part of the SNR.  Narrowband
imaging of the field could easily settle this matter, as source F
would primarily emit in emission lines if it is part of the SNR.

\section{Discussion}
\label{sec:discuss}
Using the optical data from
this paper  and the X-ray data from K00 we
can  plot the spectral energy distribution (SED) of \sgr.    We find
that any optical detections or upper limits
 severely constrain the spectrum 
below 0.5~keV (see Figure~\ref{fig:nufnu}).  The optical data given
here demonstrate that the spectrum  cannot
continue to climb when going to lower frequencies, as it 
does in the X-rays ($f_{\nu} \propto \nu^{-2.24}$ 
from 6.0~keV down to 0.5~keV; K00).   It must therefore peak
between the near UV and the hard UV/soft X-ray bands. 
The location 
of the peak depends on the form of the spectrum below the peak.  Below
we discuss several possibilities 
for the nature of \sgr, along with consequences for its optical emission.

\subsection{An Isolated NS Accreting from a Disk}
\label{sec:disk}
We modeled the emission from a fossil disk \citep{chn00,mlrh99} surrounding
a neutron 
star based on the work of \citet*{phn00} and \citet{ph00}.  Our model
accounts for both viscous dissipation  \citep{ss73} and
re-radiation of  X-rays emitted by the central source \citep{vrg+90},
and solves for the disk 
structure in a self-consistent manner.  This is similar to the
models used by \citet{hvkvk00} and \citet*{hvkk00}, and is reasonably
general.  The major assumption is that the
disks radiate like superpositions of many blackbodies.   One might think that
beaming could influence 
the results, but \citet{ph00} found that beaming from a relativistic
central object 
did little to change the observed spectrum.

\begin{figure*}
\epsscale{0.5}
\centerline{
\vbox{
\hbox{\plotone{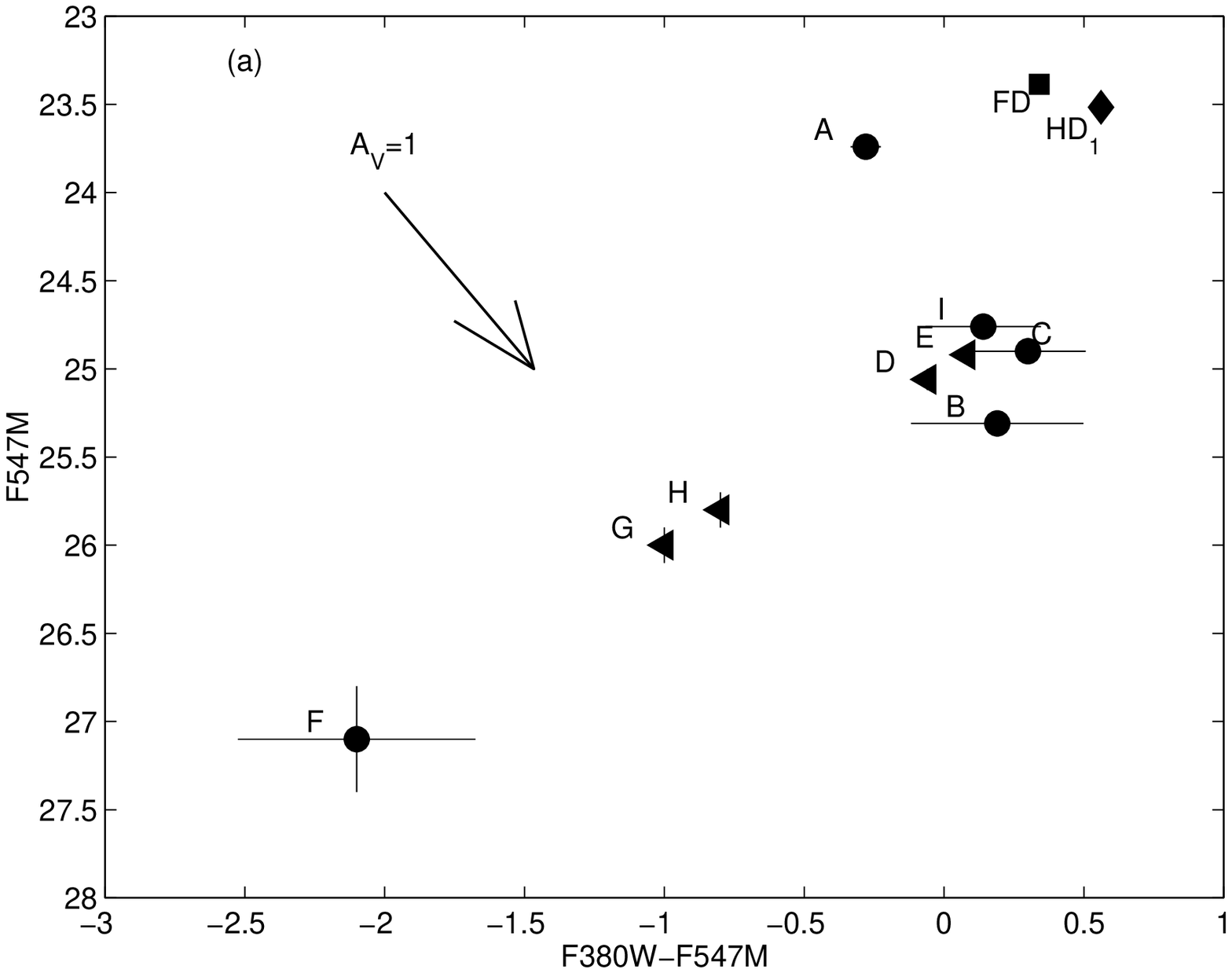}
\plotone{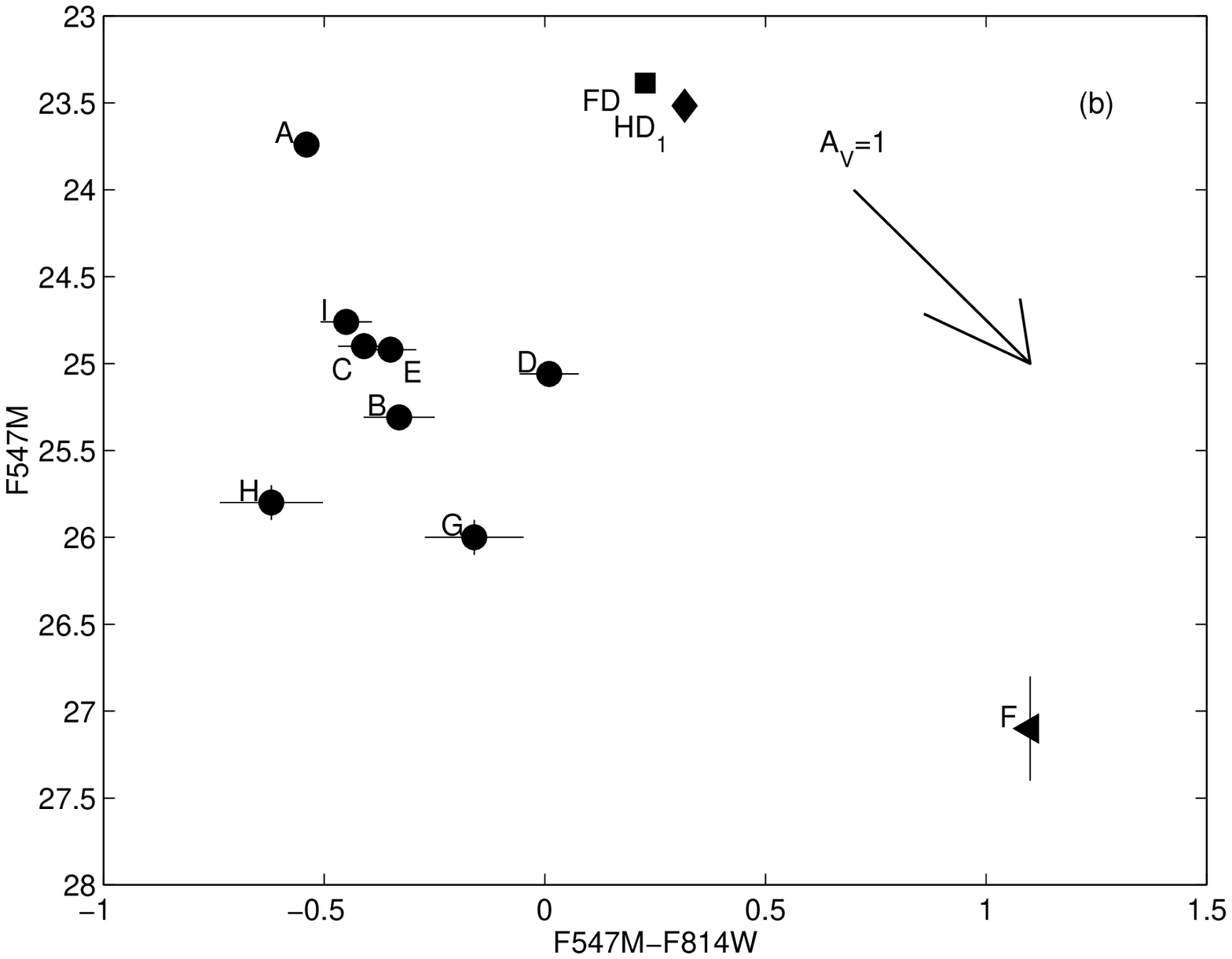}}
\hbox{\plotone{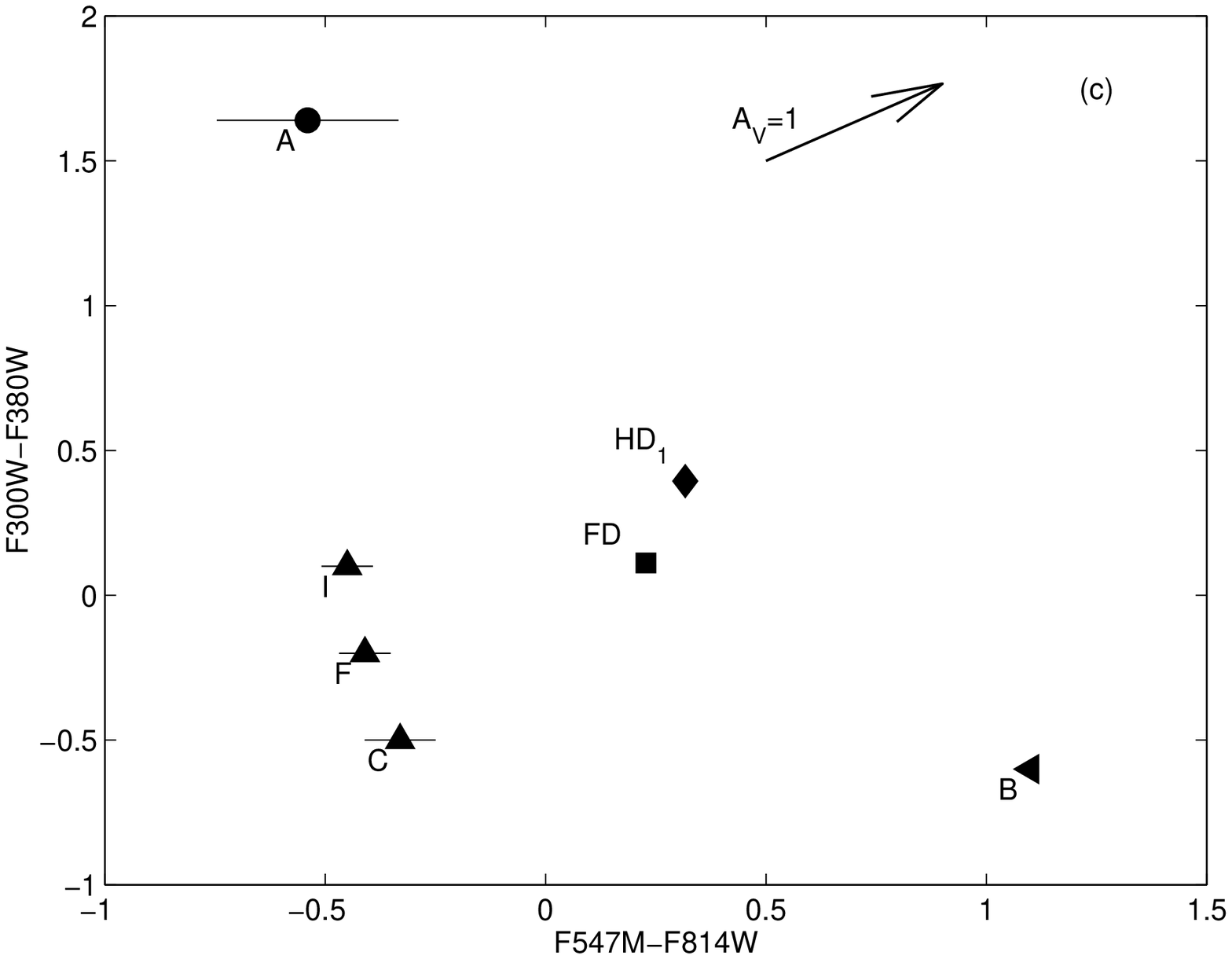}
\plotone{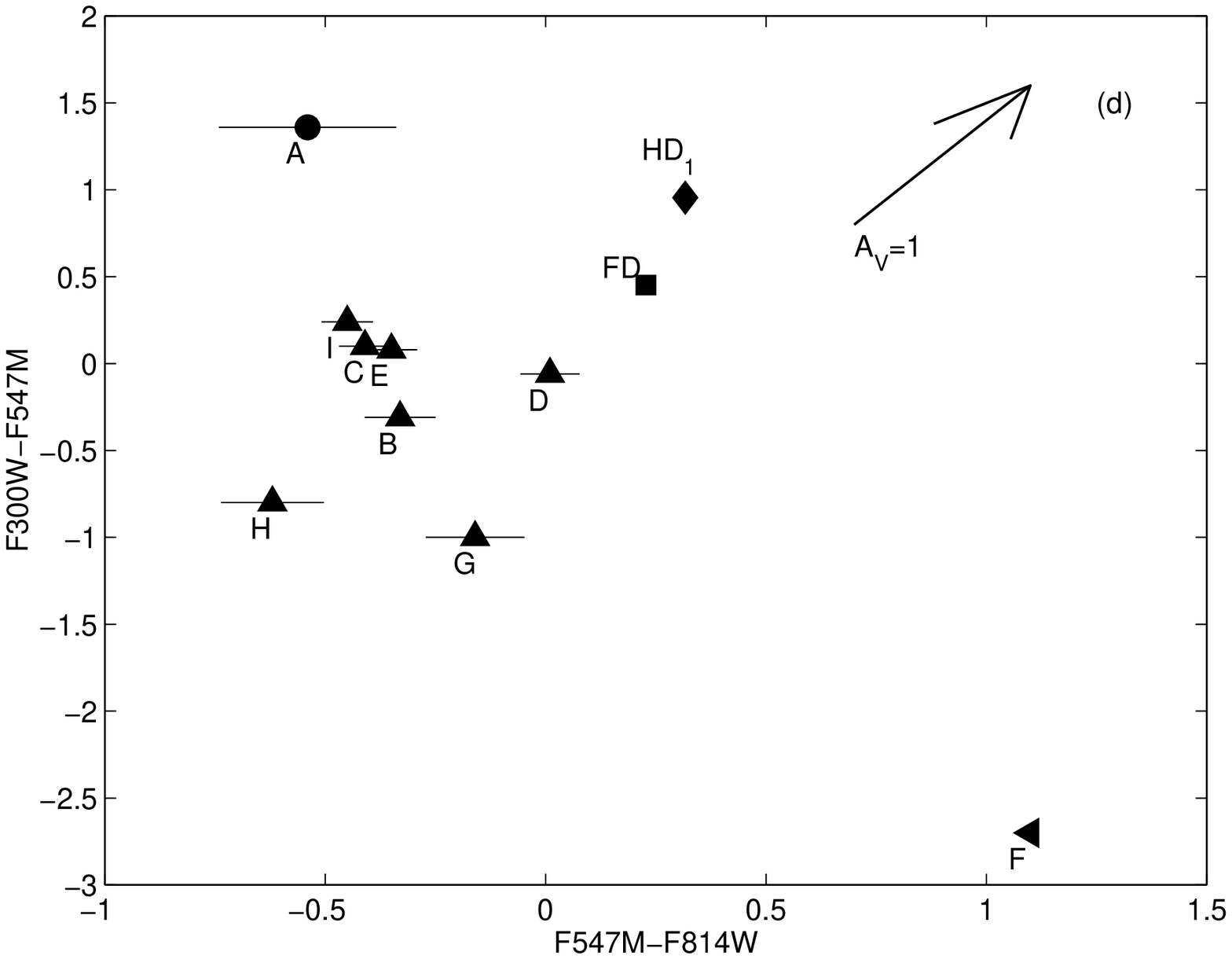}}
}}
\figcaption[figure4a.eps,figure4b.eps,figure4c.eps,figure4d.eps]{(a): F547M magnitude vs.\
F380W$-$F547M color 
for the 9 candidate sources in the \sgr\ error circle.  (b):
F547M magnitude vs.\  
F547M$-$F814M color 
for same sources.  (c): F300W$-$F380W color vs.\ F547M$-$F814W
color.  (d):  F300W$-$F547M color vs.\ F547M$-$F814W
color.  Stellar magnitudes are not corrected for
reddening.  Error-bars represent $1\sigma$.   The solid square is the
FD model (see Section~\ref{sec:disk}), while the diamond is HD$_{1}$ with
$R_{\rm in}=10R_{\rm mag}$, both of which the F547M$-$F814W
colors clearly exclude.  The disks have been reddened with
$A_{V}=1$.  Arrows indicate 1~mag reddening vectors.\label{fig:hr2}} 
\end{figure*}

\begin{figure*}
\epsscale{1.0}
\plotone{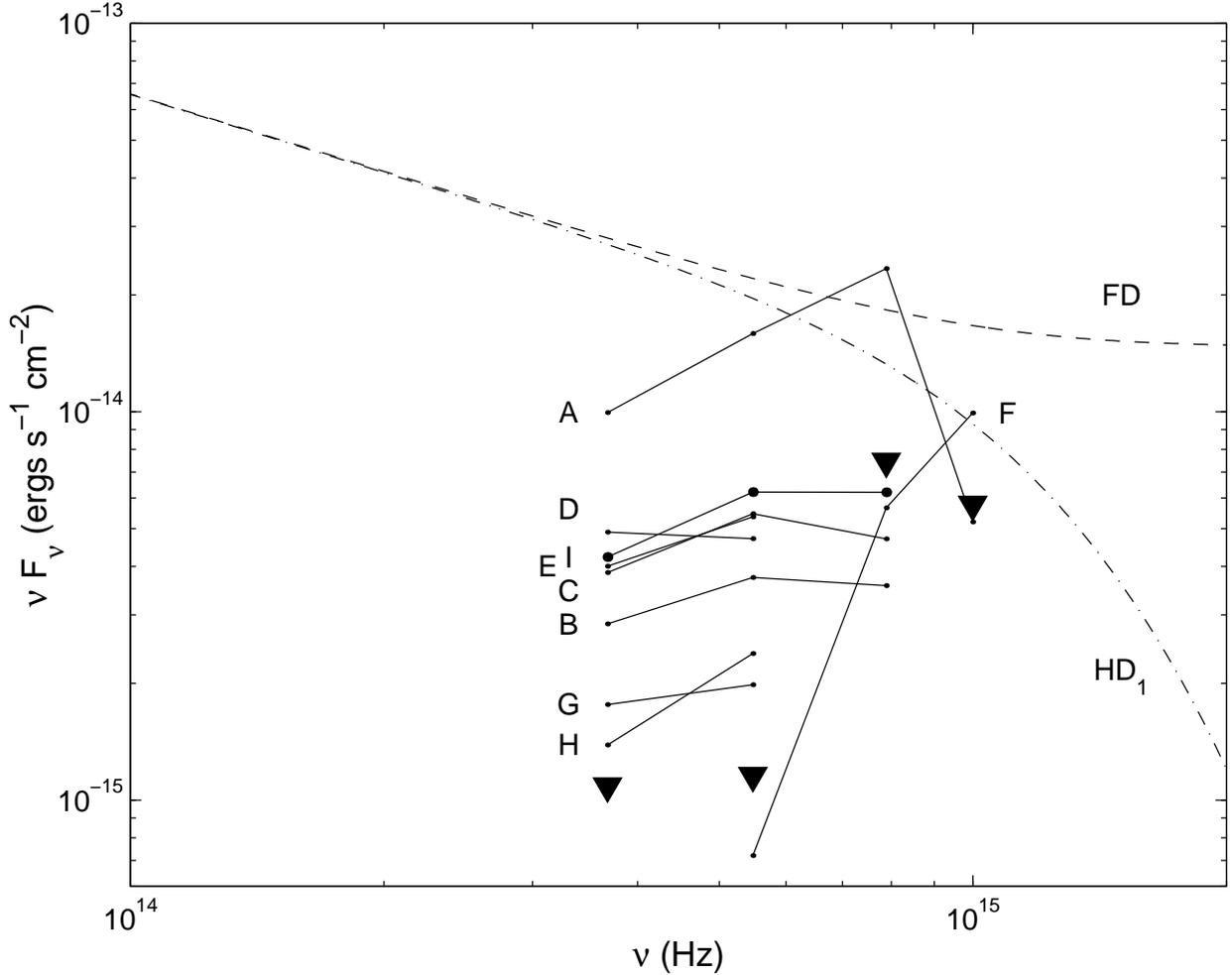}
\figcaption[figure5.eps]{Spectral energy distribution  of stars in the
\errrad\ circle 
around \sgr, compared to disk models from Perna \& Hernquist (2000).
Also see  Figure~\ref{fig:hr2}.
The individual stars are corrected for $A_{V}=1.0$, and labelled according to
Table~\ref{tab:sources}.  Also included are 3$\sigma$ upper limits for
the region around \sgr\ (upper limits on the individual stars are not plotted).
 The unabsorbed disk models are: FD (dashed); HD$_{1}$ with $R_{\rm in}=10 R_{\rm
mag}$ (dot-dashed; as suggested by Perna \& Hernquist 2000).  These models
have both the wrong fluxes and slopes compared 
to the stars.\label{fig:nufnu_stars}} 
\end{figure*}

The basic disk model (BD), plotted in Figures \ref{fig:nufnu}, \ref{fig:hr2},
and \ref{fig:nufnu_stars}, 
assumes $L_{X}=1.2 
\times 10^{36} d_{50}^{2}\mbox{ ergs s}^{-1}$ 
(0.5--10~keV; K00), which agrees with the ROSAT determination
(calculated using {\tt
W3PIMMS}\footnote{\url{http://heasarc.gsfc.nasa.gov/Tools/w3pimms.html}}
assuming $\Gamma=3.24$ and $N_{H}=8\times 10^{21}\mbox{ cm}^{-2}$).
The inner radius is the corrotation radius  
$R_{\rm in}=R_{\rm 
co}\equiv (G M P^{2}/4 \pi^2)^{1/3}=6.8\times 10^{8}$~cm ($M=1.4
M_{\sun}$, $P=8.04$~s), and the 
outer radius is $5 \times 10^{14}$~cm \citep{ph00}.  We do not
determine a value for the
outer radius independently, but the results are insensitive to it as
long as $R_{\rm out} 
\geq 10^{13}$~cm.   We also
consider an alternate model proposed by 
\citet{ph00}, where $R_{\rm in}=10 R_{\rm mag}$, which we designate HD$_{1}$
(hollow disk); here $R_{\rm mag}$
is the magnetetospheric radius defined as 
$\approx 1/2$ the
Alv\'{e}n radius $R_{\rm A}\equiv \left(\sqrt{2 G M} \dot{M} B^{-2}
R^{-6}\right)^{-2/7}$, and $B\sim 10^{13}$~G is the 
magnetic field.  As Figures~\ref{fig:hr2} and \ref{fig:nufnu_stars} show,
these models are
very likely excluded for the sources in the error circle.   The disk
models are both significantly 
brighter ($\sim 2$~mag) than the stars and have the wrong colors.
This 
result is not very sensitive to the value of the extinction, as 
our value is at the high end of reasonable values, and therefore the
corrected stellar fluxes in Figure~\ref{fig:nufnu_stars} would only decrease
upon changing $A_{V}$.  We therefore do not believe
that any of the sources in Table~\ref{tab:sources} could be an
accretion disk.  Thus the optical counterpart to \sgr\ must be
fainter than the limits presented here.

In order to accommodate
these optical 
limits for an accretion disk, we must modify 
either the inclination angle $i$, the outer radius $R_{\rm out}$, or
the inner radius $R_{\rm in}$.
For the first two parameters, acceptable values are $R_{\rm out} \leq
5 \times 10^{9}$~cm for reasonable values of  $i$ ($i\lsim 88\deg$, at
which point the central X-ray source would be hidden).  We label disks
with a truncated outer radius TD (truncated disk).  The third
parameter, $R_{\rm in}$, must increase by three orders of magnitude
above the nominal value to $\sim 10^{3} R_{\rm mag} \approx
10^{12}$~cm (HD$_{3}$ model).  This limit needs 
justification --- maintaining a disk with such an inner radius would 
be difficult.  We  plot these possible but unlikely
disks that do satisfy our limits (by changing the parameters listed
above) along with the X-ray spectrum in Figure~\ref{fig:nufnu}.

While making $R_{\rm in}$ significantly greater than $R_{\rm co}$ or
$R_{\rm mag}$ (in the HD$_{3}$ model) seems contrived, altering $i$ or
$R_{\rm out}$ are much 
less so.  For most values of $i$  we need to truncate the disk at a
radius of 
$\approx 10^{10}$~cm to accommodate our limits, similar to
\citet{hvkk00}, which as they note is difficult to produce but not
unprecedented.  It is not likely that such a truncated disk would be
able to provide 
sufficient material  to power the SGR for its lifetime.  Even so, we
examine the implications of the TD model.  

As there have been no
timing observations to constrain 
possible companions to \sgr, a companion could exist and provide an
outer limit to the disk.  
The suggested 164 day periodicity \citep{rl84} in the bursts could be
consistent with such a companion, but only if the eccentricity $e$ is
$>0.999$, which seems unlikely.

We examine the possible masses and radii of such companions in the TD
model.  We do 
this by requiring $R_{\rm out} \leq 10^{10}$~cm, where we determined
$R_{\rm out}$ as the distance from a $1.44 M_{\sun}$ neutron star to
the outermost stable orbit of a disk  (assumed to be in the orbital
plane) of pressureless particles  \citep{p77,pr80}.  This
then gives a constraint on the companion radius $R_{2}$ by requiring
that it not overflow its Roche lobe (with the Roche lobe radius
determined using  \citealt{e83}), as
this would make the source a Low-Mass X-ray Binary (LMXB) with
substantially different properties.  For a
white dwarf with $R_{\rm wd}=7.2 \times 10^{8} (M_{2}/M_{\sun})^{-1/3}$~cm
($\mu_{e}=2$), $R_{\rm wd}<R_{2}$ for all companion masses $M_{2}$ and
such a source is 
acceptable.  A main-sequence
star with $R_{\rm ms}/R_{\sun}=M_{2}/M_{\sun}$, however, would
overflow  $R_{2}$ for all $M_{2} \gsim 0.07 M_{\sun}$.  The mass limit
is even lower if
the effects of irradiation on the companion's radius \citep{p91} are
taken into account.  Also,
main-sequence stars of type K or 
earlier (and hotter white dwarfs) are excluded photometrically.  We
therefore  require a  
faint, compact  companion: a planet, brown dwarf or cool white dwarf.  
However, producing a white dwarf  
in a very tight orbit around a $10^{4}$~yr old  neutron star may be
very difficult.   

Thus the lack of an optical counterpart to \sgr\ places strong new
constraints on accretions disk models.

\subsection{A Magnetar}
\label{sec:magnetar}
There are no clear
predictions for the optical appearance of magnetars, and so one of the
stars in the error circle (Section~\ref{sec:anal})  may be such a
source.  However, based on spin  \citep{td96} and spectral (K00)
properties, we believe that the SGRs and AXPs share a common origin.
If this is the case, we can use optical data concerning AXPs
\citep{hvkvk00,hvkk00} along with the stellar appearances of sources
A--I to exclude them from consideration as  counterparts to
\sgr.  We therefore  
conclude that \sgr\ 
has not been detected.  This gives an unabsorbed X-ray to
optical flux ratio 
$f_{{\rm X}}/f_{V} \geq 3.5 \times 10^{3}$ (where $f_{\rm X}$ is the
integrated X-ray flux and $f_{V}\equiv \nu_{V}
f_{\nu,V}$).  This ratio is extremely
high --- almost all neutron stars  have lower ratios \citep{hvkk00}.  We must 
account for this in our models for \sgr.

\begin{figure*}
\epsscale{0.8}
\plotone{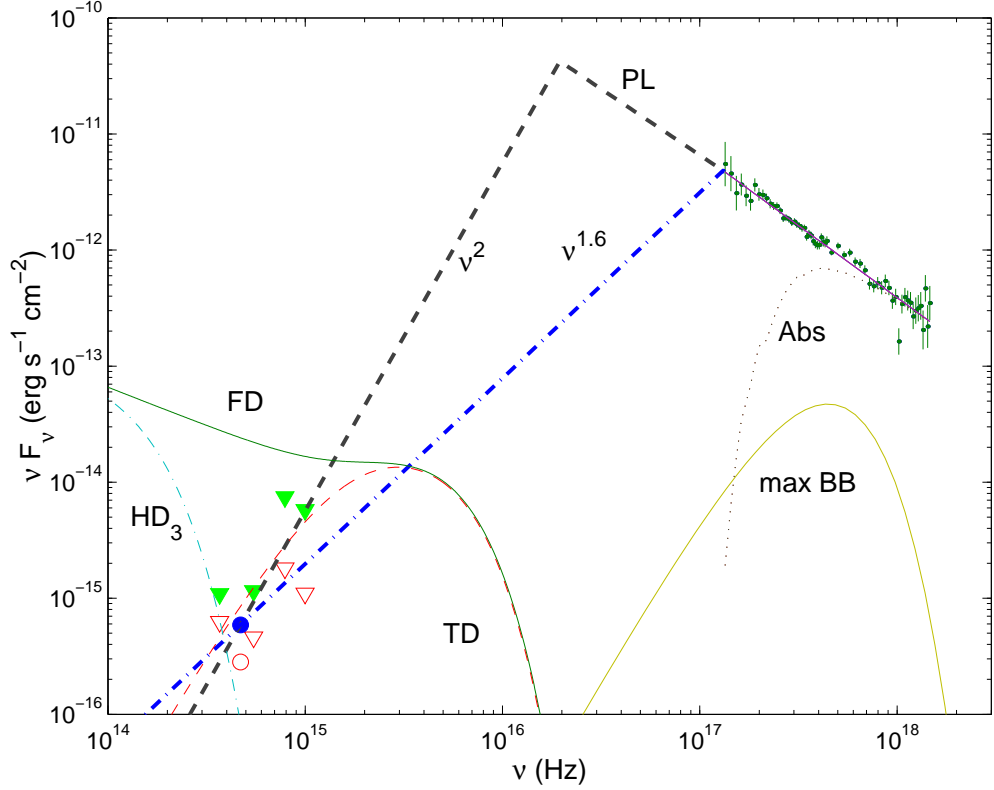}
\figcaption[figure3.eps]{Spectral energy distribution  of \sgr, including
the  X-ray
spectrum from K00 and optical limits.  Open points
denote raw 
data, while solid points are those corrected for reddening
($A_{V}=1.0$~mag).  The 
triangles are the $3\sigma$ HST upper limits, while the circles are
the estimates 
based on $f_{X}/f_{\rm opt}=7.1 \times 10^{3}$ (Hulleman, van~Kerkwijk
\& Kulkarni 2000).  The upper X-ray points are the unabsorbed fluxes,
while the lower dotted line (Abs) is the absorbed flux.   Included is the 
maximum possible blackbody contribution to the X-ray spectrum (BB).
Given the HST upper limits the spectrum of \sgr\ must peak between the
X-ray and optical bands and then decline with decreasing frequency.  We
plot two
possible power law continuations into the optical regime (PL)
with $f_{\nu} \propto \nu^{2}$ (thick dashed line) and $\nu^{1.6}$
(thick dot-dashed line), which bound likely power-law indices.  The
unabsorbed disk models are: FD (solid); TD with $R_{\rm out}=5 \times
10^{9}$~cm (dashed);  HD$_{3}$ with $R_{\rm in}=10^{3}R_{\rm mag}$
(dot-dashed).\label{fig:nufnu}} 
\end{figure*}

A continuation of the possible blackbody
emission found in the X-ray spectrum by K00 would be
extremely faint in the 
optical, and could satisfy our limits (see Figure~\ref{fig:nufnu}).
However, given the nonthermal optical emission from  radio pulsars
as well as the emission from \objectname{4U~0142$+$61} \citep{hvkk00},
we believe 
that the emission from \sgr\ would arise from another process.

We neither  know  nor can predict the details of this process, but we
 examine the emission heuristically to derive its basic properties.
A continuation of the X-ray power-law is impossible.  In general, a
 broken power-law is the simplest parameterization of the  
spectrum, with both the peak location and index below the peak
as yet undetermined.  Simple choices for the index would be 2 (thermal
emission) or 2.5 (self-absorbed synchrotron emission), although most
 indices are possible.  \citet{hvkk00} suggest that the optical emission
for \objectname{4U~0142$+$61} follows a   
power-law $f_{\nu} \propto \nu^{\alpha}$, with $\alpha \approx 2$. 
This provides the motivation for 
the lines drawn on 
the SED in Figure~\ref{fig:nufnu} extending from the optical and
meeting the X-ray spectrum in the hard-UV/soft X-ray.  The lowest
spectral index for a power-law could be 1.6, which is when the spectrum peaks
just at the lower limit of the CXO data (0.5~keV; K00), but
this seems artificial.

At the
other extreme, if the  power-law has $\alpha \gsim 2$, this would 
lead to the overall SED having a peak at $\sim 30$~nm and a
luminosity of $\sim 10^{37}\mbox{ ergs s}^{-1}$, or 10 times that
inferred from X-ray observations alone  (K00;  this also ignores particle
and the likely extremely luminous neutrino emission).  Such a
luminosity requires 
that we supply even more energy to the source during quiescence,
further constraining the energetics and leading to $B\sim 10^{16}$~G
(as noted by K00),
assuming that the source persists for $\sim 10^{4}$~yr (suggested by the
estimated age of the supernova
remnant and and the spin-down timescales of other SGRs; also see
\citealt*{cgp00}) and 
that the magnetic field is the 
primary energy supplier.  This is truly a magnetar-like  field!   The
actual case is
most likely  between the two curves on Figure~\ref{fig:nufnu}, with a
more gentle peak giving
a more relaxed energy requirement, but one that still implies $B
\gsim 10^{15}$~G.  

Such large fields
may not be consistent with the association of the SGR with N49, 
which has an age in the range $5-16$ kyr
\citep{vblr92,s83}. Calculations for AXPs \citep{cgp00}
indicate that a  
magnetar with an initial dipolar field of $B>10^{15}$~G would spin-down to  
the present day SGR spin period in only $\sim$1000~yr, and increasing
the field only makes the situation worse.  However, for \sgr\
these calculations are a simplification.  The large ($\gg 10^{13}$~G)
magnetic field would likely be 
 tangled and exist
close to the surface (K00); this would then still dominate the energetics
and produce crust stress but not the spin-down (which also must
incorporate winds; \citealt{tdw+00}; \citealt*{hck99}).  Alternatively, the
\sgr/N49 association could be false, as has recently been found for
other SGR/SNR associations \citep{hkc+99,xkl+98,lx00}.  K00 found that
\sgr\ and N49 had the same value for $N_{H}$, but they see no sign of
interaction in X-rays and the projected
position of \sgr\ is near the
edge of N49.  \sgr\ could merely be associated with nearby star
formation that led to a massive cluster (like those for SGR~1900$+$14
and SGR~1806$-$20; \citealt{fmc+99,vlh+00}; \citealt*{mfc00}), which could have produced
the N49 progenitor as well.
A non-association could then allow the age of \sgr\ to come closer to
$10^{3}$~yr, with a 
consequent reduction of $B$ by a factor of $\sim 3$.
Either way, 
the energetic requirements of the \sgr\ SED provide a strong 
constraint on the magnetar model and formation scenario for this source.

Motivated by the detection of optical emission from an AXP
\citep{hvkk00}, we examine whether it is likely that we could detect
\sgr.  Using the same  X-ray  to optical flux ratio as that for
\objectname{4U~0142$+$61} ($7.1 \times 10^{3}$; \citealt{hvkk00}),
we  estimate $R \approx 27.2$ (see Figure~\ref{fig:nufnu}), beyond
current limits but within the
range of HST.

\section{Conclusions}
We have obtained deep HST images of the field containing 
\sgr.  There does not appear to be a point-like optical
counterpart consistent with either the colors or magnitudes of the
disks described by \citet{chn00}.  Assuming a non-detection,
optical limits  place strong new constraints on both accretion disks 
and magnetars as the source of the X-ray luminosity and spindown.
Independent of the source's nature, however, if SGRs and AXPs are related, 
we might expect $R \approx 27.2$ for \sgr, based on the optical 
to X-ray flux ratio of the one identified AXP.

\acknowledgements
{This work is based on
observations with the NASA/ESA Hubble Space Telescope, obtained at the
Space Telescope Science Institute, which is operated by the
Association of Universities for Research in Astronomy, Inc. under
NASA contract No. NAS5-26555.}
We would like to thank A. Shapley for her assistance with the HST data
reduction, and R. Perna and P. Goldreich for valuable discussions.
DLK is supported by the Fannie and John Hertz Foundation, MHvK by
a fellowship from the Royal Netherlands Academy of Arts and Sciences,
and SRK by NSF and NASA.  DLK and MHvK thank the ITP at Santa Barbara,
where part of the work presented here was done, for hospitality.  The
ITP is supported by the National Science Foundation under Grant
No. PHY99-07949.

\bibliographystyle{apj}


\begin{deluxetable}{l l l r r r r}
\tablecaption{Summary of Observations \& Relevant Calibration Data\label{tab:sum}}
\tablecolumns{7}
\tablewidth{35pc}
\tablehead{
\colhead{Date} & \colhead{Filter} & \colhead{Exposure Sequence} &
\colhead {$Z_{\rm mag}$\tablenotemark{a}} & \colhead{$m_{\rm
lim}$\tablenotemark{b} 
} & \colhead{$f_{\nu,{\rm lim}}$\tablenotemark{b}} & \colhead{$A_{\lambda}$} \\
& &  \colhead{(s)} & 
&  & \colhead{($\mu$Jy)} & \colhead{(mag)}\\
}
\startdata
1998~Nov~14 & F300W & $2\times 230+8 \times 1200$\tablenotemark{c}  &
29.436 & 25.0 & 0.11 & 1.8\\
1998~Nov~14 & F380W  & $2 \times 100 + 4 \times 1000$ & 29.385 & 24.7
& 0.23 & 1.5\\
1998~Nov~14 & F547M\tablenotemark{d} & $2 \times 100 + 4 \times 1000$ & 30.695 & 26.6
& 0.084 & 1.0\\
1999~Apr~27 & F814W & $4\times 1100$ & 32.009 & 26.7 & 0.17 & 0.6 \\
\enddata
\tablenotetext{a}{STMAG zero-point.}
\tablenotetext{b}{Limiting magnitude and flux density.  See Section~\ref{sec:phot}.}
\tablenotetext{c}{This is approximate.  The actual sequence is
$200\mbox{ s}+260\mbox{ s}+4 \times 1100\mbox{ s}+4 \times 1300 \mbox{
s}$.}
\tablenotetext{d}{The F547M filter was chosen to avoid the majority of
the nebular line 
emission from N49 (\eg [\ion{O}{3}]~5007, [\ion{O}{1}]~6300,
H$\alpha$, H$\beta$), while still 
retaining spectral coverage and bandwidth.}
\end{deluxetable}

\begin{deluxetable}{c r r r r}
\tablecolumns{5}
\tablecaption{Sources in the \sgr\ Error Circle\label{tab:sources}.}
\tablewidth{32pc}
\tablehead{
\colhead{Name\tablenotemark{a}} &
\multicolumn{4}{c}{Magnitude} \\  
& \colhead{F300W} & 
\colhead{F380W} & \colhead{F547M} & \colhead{F814W} \\
}
\startdata
A & $25.1 \pm 0.2$ & $23.46\pm0.05$ & $23.74 \pm0.02$ & $24.28\pm0.02$
  \\
B & $>25$ & $25.5\pm0.3$ & $25.31\pm0.07$ & $25.64\pm0.04$   \\
C & $>25$ & $25.2 \pm 0.2$ & $24.9 \pm 0.05$ & $25.31 \pm 0.03$ \\
D & $>25$ & $>25$ & $25.06\pm0.06$ & $25.05\pm0.03$ \\
E & $>25$ & $>25$ & $24.92 \pm 0.05$ & $25.27 \pm 0.03$ 
\\
F & $24.4\pm0.1$ & $25.0\pm0.3$ & $27.1\pm0.3$ &  $\gsim 25$\tablenotemark{b} \\
G & $>25$ & $>25$ & $26.0\pm0.1$ & $26.16\pm0.05$ \\
H & $>25$ & $>25$ & $25.8\pm0.1$ & $26.42\pm0.06$\\
I & $>25$ & $24.9\pm0.2$ & $24.76\pm0.05$ & $25.21\pm0.03$ \\
\enddata
\tablenotetext{a}{See sources in Figure~\ref{fig:circle} and Section~\ref{sec:anal}.}
\tablenotetext{b}{Too extended for good photometry.}
\end{deluxetable}

\end{document}